# Traffic Engineering with Segment Routing:
# SDN-based Architectural Design and Open Source Implementation

Luca Davoli[(1)], Luca Veltri[(2)], Pier Luigi Ventre[(3)], Giuseppe Siracusano[(3)], Stefano Salsano[(3)]
(1) Univ. of Parma / Consortium GARR - (2) Univ. of Parma - (3) Univ. of Rome Tor Vergata

*Abstract* – Traffic Engineering (TE) in IP carrier networks is one of the functions that can benefit from the Software Defined Networking paradigm. By logically centralizing the control of the network, it is possible to "program" per-flow routing based on TE goals. Traditional per-flow routing requires a direct interaction between the SDN controller and each node that is involved in the traffic paths. Depending on the granularity and on the temporal properties of the flows, this can lead to scalability issues for the amount of routing state that needs to be maintained in core network nodes and for the required configuration traffic. On the other hand, Segment Routing (SR) is an emerging approach to routing that may simplify the route enforcement delegating all the configuration and per-flow state at the border of the network. In this work we propose an architecture that integrates the SDN paradigm with SR-based TE, for which we have provided an open source reference implementation. We have designed and implemented a simple TE/SR heuristic for flow allocation and we show and discuss experimental results.

## I. Introduction

The Segment Routing (SR) paradigm [2] aims at providing enhanced packet forwarding behavior without requiring per-flow state maintenance within the network, leading to a reduction of the complexity for both control and user planes. With SR the path of a packet can be enforced through an ordered list of processing/forwarding functions, called segments, that may consist of both logical and physical elements: for example a segment may be a packet filter, a network node, or a network link. The chain of these elements forms the *SR path* of a packet, identified by a list of segment identifiers (SIDs). The scope of such SIDs can be global or local. While global SIDs are defined globally and should be recognized by all network nodes, local SIDs are defined locally within a node and the use of local SIDs by other nodes requires the implementation of an explicit distribution mechanism.

SR leverages on the source routing concept. The list of SIDs representing a SR path is inserted within the packet. If the list of SIDs is inserted only at the border of a SR domain, per-flow state is required to be maintained only by the ingress nodes, while the internal nodes will forward the packets only based on the SIDs carried by these packets: no per-flow state should be maintained and no node reconfiguration is required.

SR is a general concept that needs to be mapped onto a specific forwarding technology that supports source routing, for example MPLS or IPv6. In this paper we restrict our attention to MPLS-based Segment Routing, in which SIDs are realized with MPLS labels.

Traffic Engineering (TE) is an example application of SR. In the context of a TE application, SR allows the set-up, modification, and tear-down of *TE paths* within a network domain, operating only at the border of the network. This approach minimizes the state information needed within the network (since only ingress/egress nodes should maintain per-flow state) and the rate of table update requests.

The control plane for SR can be managed either in a distributed or in a centralized way. We focus on a centralized approach, combining SR with Software Defined Networking (SDN) [3]. The basic concept of SDN is the decoupling of the network control and forwarding functions, enabling the underlying infrastructure to be abstracted and programmable in the control plane. The network control function is logically centralized in an entity called *controller* that provides an abstract and centralized view of the overall network to the SDN applications running on top of the control plane. The interface between the application plane and the control plane is called "NorthBound Interface" (NBI), while the interface between the control plane and the data plane is called "SouthBound Interface" (SBI). OpenFlow [4] is a protocol that has been specifically designed for the SBI.

SDN solutions can be applied in different scenarios, from enterprise data centers to corporate/campus networks, till large carrier networks. In some cases a pure SDN solution based on Layer-2 switches interconnected to a centralized controller (e.g. in data centers) can be used, while in other contexts a more complex solution including the interaction with standard Layer-3 routing protocols is needed. We have recently proposed the Open Source Hybrid IP/SDN (OSHI) [5] framework. The OSHI data plane is based on a hybrid approach allowing the coexistence of traditional IP routing with SDN-based forwarding within the same provider domain. Such an approach has been defined in [6] as "Service-Based" or "Class-Based" Hybrid SDN (depending on how the IP- and SDN-based services are combined). In [7] we describe the design and implementation of two basic services in the OSHI framework that will be reused in this work: the IP "Virtual Leased Line" (IP VLL) and the Layer 2 "Pseudo-wire" (PW). The former is used to interconnect two remote IP end-points with a virtual point-to-point link. The latter provides a fully transparent virtual Ethernet interconnection between two remote end-points. In the OSHI architecture, the IP VLL and PW services are realized using MPLS labels for tunneling and forwarding, while all the control plane is based on SDN/OpenFlow (no traditional MPLS control plane is used). Further details are reported in [7].

The main contributions of this paper are: i) design and implementation of data plane and SDN-based control plane for Segment Routing based on OSHI; ii) design and implementation of SR path assignment algorithm for TE (section III); iii) a preliminary performance analysis of the proposed solution. All the developed code is open source and available at [8]. For an easier reproducibility of the experiments, we have packaged all software components in a ready-to-go virtual machine (available at [5]).

## II. REFERENCE ARCHITECTURE

We consider an ISP network managed by a (logically) centralized SDN controller, as shown in Figure 1. Network nodes are classified in Provider Edge (PE) routers and Core Routers (CR), both types are assumed to be MPLS nodes. As in a traditional MPLS network, PE nodes are capable to originate and terminate connections (Label Switched Paths), while both PE and CR nodes are capable of switching labels, i.e. they can be in the middle of a path. The PE nodes are connected to the Customer Edge (CE) routers, which are the external traffic sources and destinations.

The PE and CR nodes are *hybrid* IP/SDN nodes according to the OSHI architecture proposed in [5]. According to the OSHI architecture, no MPLS control plane needs to be implemented within these nodes: a SDN approach is used to program the flow tables of the switching component in a node, here denoted as OFCS (OpenFlow Capable Switch), to stress its capability to be programmed via the OpenFlow (OF) protocol.

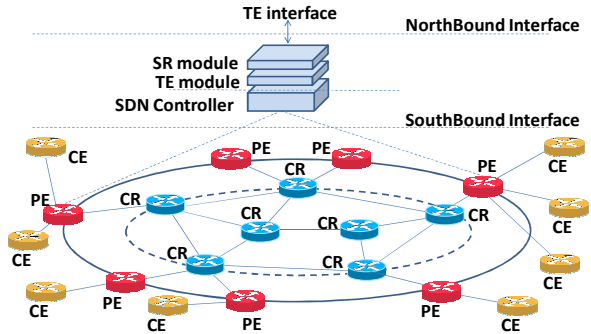

Figure 1 - Traffic Engineering in a SDN-enabled network.

We implemented Segment Routing on top of the MPLS functionality provided in [5]. The flow tables are now managed in combination by the external SDN controller and by a local entity called "SR daemon", according to the node architecture depicted in Figure 2.

The SDN controller is in charge of setting up the *edge-to-edge* services, by configuring the ingress and egress PE nodes to support a given flow (section IV.B). Thanks to the SR approach, no configuration of internal nodes is needed to support a specific flow.

The local SR daemon is in charge of configuring the flow tables to support the MPLS labels representing the SIDs. The SR daemon locally interacts with the routing daemon for obtaining routing table updates and consequently programming the switch flow tables (details in section IV.A). In this way, the SR solution relies on the functionality of the interior routing protocol (OSPF in our case), but as a requirement we want to minimize the enhancements of the routing protocol to support SR.

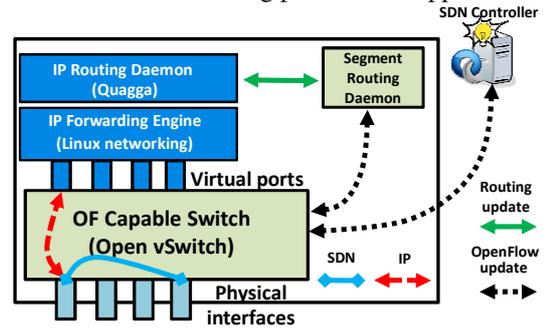

Figure 2 – High level architecture of a SR enabled OSHI node.

As a further requirement, we want that SR-based forwarding can coexist with traditional hop-by-hop MPLS-based Label Switched Paths (LSP). This way, we can offer IP VLL and PW services based either on hop-by-hop LSP or on SR forwarding.

## III. TRAFFIC ENGINEERING FOR SEGMENT ROUTING

With reference to Figure 1, we enhance a SDN controller with TE/SR modules. The TE/SR features could be offered to applications through a proper NorthBound Interface, however the specification of such NBI is out of scope of this paper. On the SouthBound Interface, we considered the OpenFlow protocol for our demo implementation. We assume that the SDN controller is requested to allocate a set of traffic flows with a specified bit rate, knowing the link capacity. The TE/SR modules will first allocate hop-by-hop *TE paths* solving a classical *flow assignment* problem (section III.A). Then for each TE path it will compute a corresponding *SR path* (section III.B, *SR assignment* problem) for instructing the flow packets through the assigned TE path.

Our assumption is to execute flow assignment and SR assignment algorithm in sequence without interaction. A combined procedure could achieve further optimization of some parameters, but it is out of the scope of this paper.

### A. Basic TE algorithm

We consider a classical flow assignment problem: given a set of (unidirectional) flows between nodes with their expected rate (b/s) and given the link capacities (b/s), find the "optimal" paths for all the flows. We assume that there can be more than one flow between the same (source, destination) node pair, and that each of those flows can be individually routed through a different hop-by-hop path, without the possibility to split a single flow into multiple paths. We include a feasibility/admission control check: a solution is admissible only if on each link the sum of the rates of the allocated flows does not exceed the link capacity. Therefore the "optimal" solution could also include only a subset of the input flows.

For the optimality criteria, we follow the approach identified in [11] and [12]. We consider an hypothetical network scenario in which the traffic flows are characterized with Poisson arrivals, distribution of packet length exponentially negative, and independence between arrivals and departures. In these conditions each link can be modeled as M/M/1 queue, and the queueing time can be easily evaluated knowing the load on the link. For a given allocation of flow paths, it is therefore possible to evaluate the global average crossing time $T_{avg}$ experienced by packets. In fact, thanks to the Jackson result [13] the average delay on a path can be expressed as the sum of the delays encountered on the links of the path and then a weighted sum based on the flow rate is used to evaluate the global average crossing time. We define the flow allocation as an optimization problem trying to minimize $T_{avg}$. This is a NP-complete problem that we face with the heuristic proposed in [11] and [12]. Even if the hypothesis on the traffic flows are not realistic, this approach is very effective in equalizing the load on links and tries to spread traffic evenly on the network, avoiding critical bottlenecks.

The input of the heuristic consists in the topology, the capacity of the links, and the traffic flows, represented as triples (source, destination, bit rate). The heuristic is divided in two phases: i) a Constrained Shortest Path First (CSPF) phase, where a first allocation of the flows is realized (flows are rejected if they cannot be allocated in this phase); ii) a heuristic re-assignment phase, which tries to re-assign all admitted flows one-by-one, in order to minimize the global network crossing time $T_{avg}$. The second phase is executed multiple times until no improvement is achieved. The details of the procedure can be found in [14]. Note that in this paper we are not focused on the merits of this heuristic, but we only need a TE algorithm that allocates flows into hop-by-hop *TE paths* to be transformed into *SR paths*.

*B. Mapping TE paths into Segment Routing paths*

In order to enforce TE through SR, the SR assignment algorithm is performed on the hop-by-hop TE paths obtained from the heuristic.

For example, let us consider the topology of Figure 3 and the TE path $\{n_1, n_2, n_3, n_5, n_6\}$ indicated with thick arrows. We are assuming that there is a single IP link between two nodes, so that the sequence of nodes univocally identifies the path. We want to find a SR path, i.e. a list of segments that enforces the same route in the network. A suitable SR path is $\{n_1, n_3, n_5, n_6\}$, where $n_3, n_5, n_6$ are *node segments* or *node SIDs*. The node segments are global SIDs which simply instruct an intermediate node to send the packet following the shortest path towards the node corresponding to the SID.

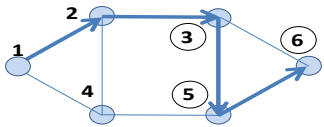

Figure 3 – Example: network topology and TE/SR path

In the above example, a sequence of node segments was able to represent the given TE paths, but this is not always the case. Let us consider the topology represented in Figure 4 and the TE path $\{n_1, n_3, n_5, n_7\}$. The SR path $\{n_1, n_3, n_5, n_6\}$ is not able to enforce the same route: node $n_3$ will find equal cost paths towards $n_5$ over the links $n_3{\rightarrow}n_4$ and $n_3{\rightarrow}n_5$ and node $n_5$ will send the packet towards $n_6$ on the link $n_5{\rightarrow}n_6$ instead of using the link $n_5{\rightarrow}n_7$. In the Segment Routing architecture, this issue is solved with the use of IGP-Adjacency Segment or "Adj-SID". This segments needs to be advertised by the nodes and can be of local or global significance. For example, let us respectively represent with SIDs $l_{3>5}$ and $l_{5>7}$ the segments corresponding to the direct links $n_3{\rightarrow}n_5$ and $n_5{\rightarrow}n_7$. Using Adj-SIDs with local significance, a suitable SR path is: $\{n_1, n_3, l_{3>5}, l_{5>7}\}$. In principle, it is also possible to use Adj-SIDs with global significance, leading to the following shorter SR path: $\{n_1, l_{3>5}, l_{5>7}\}$. The price is the need to install one additional rule in all nodes for each global Adj-SID to enforce the proper routing.

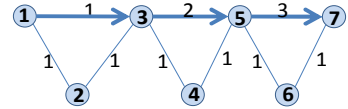

Figure 4 – Network topology in which SR node segments are not able to enforce arbitrary TE paths (one example shown in the figure).

The use of Adj-SIDs is not desirable for the following reasons: the dissemination of Adj-SIDs introduces complexity and is based on enhancements to IGP protocols; if local significance SIDs are used, the handling of node/link failures is more complex or less efficient; if global significance SIDs are used, the state information in nodes grows and the scalability decreases. For this reasons, we propose a solution with a new type of SID that has global significance but does not need to be disseminated, as it can be automatically derived and handled by all nodes. We call this type of Segment ID as *direct-link* SID and represent them as $n_k^*$. A *direct-link* SID instructs a node to send the packet on the direct link toward a target node, if the direct links exists. Otherwise the regular shortest path forwarding towards the target node is used. The *direct-link* SID $n_k^*$ has global significance and, in particular, it can be used by different neighbors of a node $n_k$ to send the packet on the direct link towards $n_k$, even if this link is not the shortest path toward $n_k$.

Note that there are topological conditions under which only node SIDs are enough to enforce any TE path, and the use of *direct-link* SIDs is not needed. In particular, we should assume that when a direct link exists between two nodes, it matches the shortest path between those nodes. This is easily verified in case all links have the same cost, which constitutes a sufficient condition for not having to use the *direct-link* SIDs.

The objective of the SR assignment algorithm is to find the minimal-length SR paths corresponding to the TE paths, i.e. the shortest list of SIDs that allows the packets to follow the assigned TE path.

In the following, the node SID associated to a node $n_k$ will be simply referred through the name of the node $n_k$, while the *direct-link* SID for $n_k$ is represented with $n_k^*$. Let us suppose that a given flow *f*, characterized by the ingress PE node *I* and the egress PE node *E*, is requested to be setup with an assigned hop-by-hop TE path, specified by the complete list of all intermediate nodes $R_1$, $R_2, .. , R_{N-1}$.

The complete hop-by-hop TE path is:

$$P = \{ R_0=I, R_1, R_2, .. , R_{N-1}, R_N=E \}$$

The *SR assignment* problem consists in finding a sequence of SIDs that can be pushed into a packet by the ingress PE node *I*, forcing the packet to follow a given route; this sequence of SIDs is the SR path of the packet. We propose a *SR assignment* algorithm, using the following notation:

- *G*: the graph of the network;
- *P*: the assigned TE path;
- $tep(n_s,n_d)$: portion of the TE path starting from node $n_s$ and ending with node $n_d$. As particular case, $tep(I,E) \equiv P$;
- $SP(n_s,n_d) = \{ sp_i(n_s,n_d) , i=1,2,..,M \}$: the set of equal-cost shortest paths $sp_i(n_s,n_d)$ in *G* from $n_s$ to $n_d$, based on the current routing tables; routing tables are considered to be already set-up using a link-state routing protocol (e.g. OSPF), using Shortest Path First algorithm;
- $SP^*(n_s,n_d)=\{ sp_i^*(n_s,n_d) , i=1,2,..,M^* \}$: the set of *direct-links biased* equal-cost shortest paths in *G* from $n_s$ to $n_d$;
- $prec(p,n)$: the preceding node of node *n* along a path *p*;
- *srp*: the SR path containing the list of assigned SIDs.

The pseudo-code representation of the *SR assignment algorithm* is reported in Figure 5. The algorithm takes as input the graph of the topology and the assigned TE path, and returns as output the assigned SR path (the segment list). It starts by considering as current source node $n_s$ the ingress node *I* of the TE path and as current target node $n_d$ the egress node *E* of the TE path. Then the set of equal-cost shortest paths from $n_s$ to $n_d$ is considered. If only one shortest path exists and it equals the TE path (from $n_s$ to $n_d$), then the node $n_d$ is used as node SID from $n_s$ to $n_d$, and the algorithm ends. On the contrary, if the number of equal-cost shortest paths is greater than 1 or the shortest path differs from the TE path, the set of *direct-links biased* equal-cost shortest paths is considered. If the size of the set is 1 and the only path equals the TE path (from $n_s$ to $n_d$), then the node $n_d$ is used as *direct-link* SID from $n_s$ to $n_d$ and the algorithm ends. On the contrary, if the two paths differ or the size of the set is greater than 1, no single SID exists for the entire path from $n_s$ to $n_d$, and the previous procedure is repeated considering as target node the preceding node of $n_d$ (that is new $n_d=prec(p,n_d)$). If it succeeds in finding a SID, then the SID is added to the segment list and the procedure is repeated from the current $n_d$ to *E* (that is $n_s=n_d$, $n_d=E$); otherwise, if it fails, $n_d$ is replaced with the preceding node (as above), and the procedure is repeated. The algorithm ends when a SID is found with $n_d=E$.

For example, let us consider the topology of Figure 3, and suppose that we want to find a SR path for the TE path $p=tep(n_1,n_6)=\{n_1,n_2,n_3,n_5,n_6\}$. Since there are more than one shortest between $n_1$ and $n_6$, the sub-path $p=tep(n_1,n_5)=\{n_1,n_2,n_3,n_5\}$ is considered. Since the shortest path between $n_1$ and $n_5$, that is $sp(n_1,n)=\{n_1,n_4,n_5\}$, differs from *p*, the sub-path $p=tep(n_1,n_3)=\{n_1,n_2,n_3\}$ is considered. Now, since *p* coincides with the shortest path $sp(n_1,n_3)$, $n_3$ is added to the SR path and the algorithm restarts between $n_3$ and $n_6$. At the end the returned SR path will be $srp=\{n_1,n_3,n_5,n_6\}$. In this example, all the steps can be enforced with node SIDs (as it was expected, considering that all links have the same unitary cost).

In the implementation, we automatically derive both the global node SID and the *direct-link* SID from the loopback interface address that can univocally identify a (MPLS) router in an ISP network, as described in section IV.C. This is an important simplification, as there is no need to distribute SIDs through extensions of the routing protocols.

```
function SEGMENTALLOCATION (G , P) --> srp
    n_s = I,  n_d = E,  srp = { }
    BEGIN:
        p = tep(n_s, n_d)
        // check if p is equal to the (only one) shortest path
        if ((sizeof(SP(n_s,n_d)) == 1) AND (sp_1(n_s,n_d) == p))
        then
            ADD n_d to srp
            if n_d == E
            then
                goto FINISH:
            else
                n_s = n_d ;  n_d = E
                goto BEGIN:
        else
            // p is not the (only one) shortest path,
            // try with direct-links
            if ((sizeof(SP*(n_s,n_d)) == 1) AND (sp_1*(n_s,n_d) == p))
            then
                ADD n_d* to srp
                if n_d == E
                then
                    goto FINISH:
                else
                    n_s = n_d ;  n_d = E
                    goto BEGIN:
            else
                // p is not the (only) direct-link shortest-path,
                // try with a shorter segment
                n_d = prec(p, n_d)
                goto BEGIN:
    FINISH:
```

Figure 5 – Second phase of the algorithm

IV. OPEN SOURCE IMPLEMENTATION OF THE DATA AND CONTROL PLANES FOR MPLS-BASED SEGMENT ROUTING

*A. SR daemon*

The SR daemon manages the flow tables of the local OFCS, providing the initial configuration and inserting the information coming from IP routing. The flow tables needs to be configured to handle node SIDs and *direct-link* SIDs coded in MPLS labels. The SIDs can represent the node itself or a remote node. Let us first consider how the node handles the SIDs that represents the node itself (one node SID and one *direct-link* SID). When the node starts, the SR daemon initializes the OF tables to handle these SIDs with this rule: pop out the outer MPLS label and resubmit to the OF tables so that the next label is processed. All the other SIDs representing the remote nodes are managed after this initialization phase, when the

SR daemon interacts with IP routing using the "FIB push interface" [10] offered by Quagga, based on the so-called Forwarding Plane Manager (FPM) protocol. The SR daemon listens for connection on the well-defined FPM TCP port (2620) and then receives the update messages from Quagga. When the SR daemon receives a *route update* message for a loopback address (i.e. the identifier of one of the PE or CR nodes), it starts an *add-SID-destination* procedure for the node SID. During this procedure, the daemon generates the MPLS labels of the SR-based IP VLL and PW services from the loopback address, then it retrieves the MAC address of the next-hop node, finally it installs the needed rules in the OF table. The *direct-link* SID is also added with the following approach. If the node has a direct link for the destination node, a rule is added that forwards over the direct link. If the node does not have a direct link towards the destination node, a rule is added using the same output link derived from the routing protocol. When the SR daemon receives a *route delete* message for a loopback address, it will delete all the rules added for this destination.

### B. Remote programming of the OF tables

A SR path is characterized by the input port in the ingress PE, the list of segments (i.e. MPLS nodes to be crossed) and the output port in the egress PE. The SR path is unidirectional: in case of bidirectional paths, the lists of segments in the two directions do not need to be symmetric. The following operations are needed to setup a unidirectional SR path; we note that they will act *only* on the ingress and egress PE, as no operation is needed in the core to install the SR path.

1. Allocate the MPLS label associated to the destination output port toward the CEs, as needed to perform the egress operation in the PE (it will be the inmost label);
2. Evaluate the MPLS labels associated to the nodes in the list of segments;
3. In the ingress PE: insert an OF rule that pushes first the MPLS label associated to the output CE and then the MPLS labels of the crossed nodes in the correct order;
4. In the egress PE: insert the rule that strips the last MPLS label (the one associated to the CE) and forwards on the port connected to the output CE.

As a result, the ingress PE will add the proper stack of MPLS labels to the packets coming from the CE, the internal nodes will forward using traditional MPLS forwarding, and the egress PE will strip the inmost label and forward a plain IP packet toward the destination CE. At time of writing, there is a maximum number of 4 MPLS labels in a packet that can be handled by Open vSwitch (OvS). We have realized a custom version of OvS where this limit has been increased to 6, meaning that we can leverage on 5 labels to steer the traffic in the network.

We have implemented these operations using a Python script on top of the Ryu controller. The script is called VLL Pusher and extends the one we described in [7].

### C. Allocation of the MPLS label space

To allocate the MPLS label space, we considered these requirements: i) to support both hop-by-hop MPLS LSP and SR-based paths; ii) to support both IP VLL and PW services. The MPLS label is 20 bit long and we use the leftmost bits to identify different service types (Table 1).

| 000 | Hop-by-hop LSP for IP VLL service |
|---|---|
| 001 | Hop-by-hop LSP for PW service |
| 010 | Endpoint of a SR-based IP VLL |
| 011 | Endpoint of a SR-based PW |
| 10000 | SR-based path for IP VLL service (node SID) |
| 10001 | SR-based path for IP VLL service (dir. link SID) |
| 10010 | SR-based path for PW service (node SID) |
| 10011 | SR-based path for PW service (dir. link SID) |
| 101 | OAM traffic |
| 110, 111 | Currently unused |

Table 1 – Allocation of MPLS label space.

With the above assignment, we use 17 bits to identify the endpoint of a SR path (e.g. the output port of the egress PE corresponding to the destination CE). We use 15 bits to identify the different MPLS nodes, extracting the 15 rightmost bits from the node loopback address distributed using OSPF, and we use one bit to differentiate between node SIDs and *direct-link* SIDs for the same target node.

### D. Summary of design choices

Here we summarize the main design choices for our implementation for MPLS-based Segment Routing: i) MPLS labels are used as segments; ii) local segments are not currently supported; iii) the 16 rightmost bits of the nodes loopback addresses are encoded in the label; iv) there is a common MAC address for all interfaces in an OSHI node (like in a switch) and a static mapping exists between SR nodes and their MAC addresses, globally known from each node; v) the inmost level of the MPLS stack is used to identify a specific CE, thus it is exploited to forward the packets towards the correct access interface; vi) the pop action of the inmost MPLS label happens only in the destination node and not in the penultimate hop; vii) the range of MPLS labels is divided in classes (see section IV.C).

## V. TE/SR IMPLEMENTATION

The proposed architecture and the algorithms described in section III have been implemented and tested. Figure 6 shows the developed software components and their interactions. The source code is available at [8]. The topology parser can acquire topology information from the REST API of Ryu controller, from the GraphML representation used in the Topology Zoo project [15], or from our GUI (Topology 3D GUI [7]). A random demand generator can be used to create a list of traffic flows between PE nodes (or the list of flows can be manually created with the Topology 3D GUI). Two Java modules implement the flow assignment heuristic and the SR assignment algorithm. Their output is a list of SR paths. Both the input to the flow assignment heuristic and the output of the SR assignment algorithm are represented as JSON files (examples in [8]). The SR paths list is processed by the *OSHI_SR_pusher* python script. Using the Ryu REST API, this script configures the ingress and egress operations in PE OSHI nodes, as discussed in section IV.B. The data plane is emulated in Mininet, with OSHI nodes extended by the OSHI-SR-dataplane - i.e. the extension module implementing the SR daemon described in section IV.A.

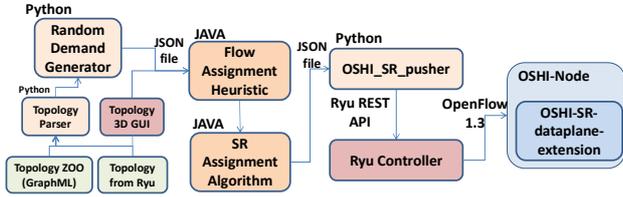

Figure 6 – Software components of the implementation

## VI. EVALUATION OF THE SOLUTION

Based on our implementation, we carried out an experimental analysis of the proposed architecture and algorithms, with the following two main goals:
- testing the SR assignment algorithm;
- testing the overall implementation of the solution, from the SDN-based control plane to the MPLS-based SR forwarding in the data plane.

To accomplish the first goal, we considered a relatively large scale topology with 153 nodes and 354 unidirectional links, the "Colt Telecom" topology included in the Topology zoo dataset [15]; we assumed that all links have the same capacity. We developed a random demand generator as follows. We randomly selected 40% of the nodes to be PE (e.g. ingress/egress), then we randomly selected 20% of the PE couples to be active source/destination of traffic flows. For each active couple of PEs, in each direction we have an average of 3.5 flows (the number of flows has a geometrical distribution) with the sum of the flow rates equal to 10% of the capacity of a link and the size of each flow has a negative exponential distribution. With these parameters we generated a list of 2460 flows with their bit rate. As reported in Figure 8, we have run a set of allocation experiments by selecting an increasing number of flows out of the full demand of 2460 flows and we have collected the number of allocated flows (the flows are selected following the same order). The straight line show the ideal case with infinite link capacity.

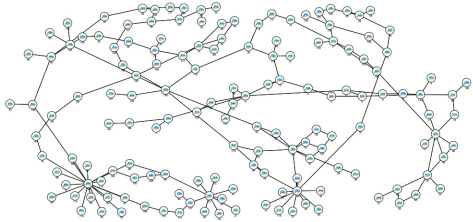

Figure 7 – Colt Telecom (08/2010) topology from Topology Zoo (each link in the picture corresponds to two unidirectional links)

This demand has been generated to largely overcome the network capacity, so that only a subset of the flows can be allocated. In this condition we stress the flow allocation algorithm and we can have TE paths that diverge from the shortest path.

In the first experiment, the full set of requested flows (2460) is considered, resulting in the allocation of 940 flows. For each allocated flow, the TE path and SR path have been computed according to the flow assignment heuristic and SR assignment algorithm. Figure 9 reports the distribution of path lengths for the TE paths and for the *natural paths*. With natural path of a flow we refer to the shortest path from ingress PE to the egress PE (with no capacity constraints). As expected, TE paths are longer than natural paths (in this experiment the mean length is 8.17 vs 6.63).

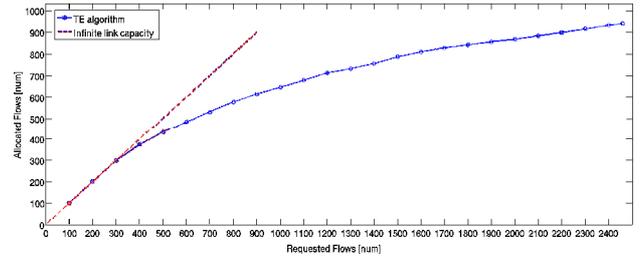

Figure 8 – Allocated flows vs. requested flows.

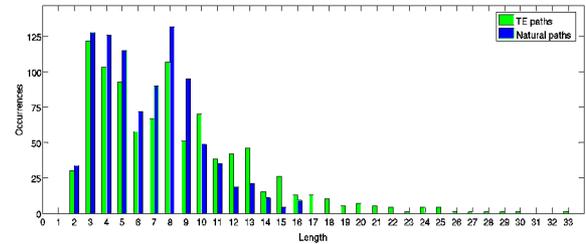

Figure 9 – Distribution of path length for TE and natural paths

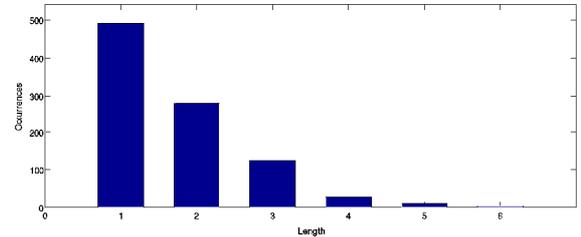

Figure 10 – Distribution of SR path length (number of SIDs in the path)

The result of the SR assignment algorithm is shown in Figure 10, which provides the distribution of SR path lengths. It is interesting to note that for more than 50% of the flows the SR path has only 1 SID, i.e. the egress node (this means that the flow is surely allocated on a natural path). When the SR path has more than one SID, it can be a natural path in which it was needed to differentiate among multiple equal cost shortest path, or a TE path different from the natural path. It is interesting to note that the mean number of hops in a SR path is low: ~1.7 and the maximum number is 6.

The flow assignment and SR assignment algorithms have been also tested varying the number of allocated flows. In Figure 11 we show the mean length of TE paths, natural paths, and SR paths for different number of allocated flows. Increasing the number of flows (i.e. the traffic load), the mean length of the TE paths increases from ~8.2 to ~9.7. When the number of allocated flows further increases, the mean length starts to decrease. This depends on how we have selected the sets of allocated flows starting from the full list of requested flows (2460).

As shown in Figure 8, we extract larger subsets from the full list and select the admitted flows from this subset. This creates a bias in the selected flows: when we have already admitted several flows and loaded the network, the flows with distant ingress and egress PEs can hardly be allocated, therefore the flows that are accepted are on average shorter. The mean number of SR-hops starts from ~1.3 when the network is not loaded and TE paths correspond to natural paths, because of the segments used to differentiate among multiple equal-cost shortest paths.

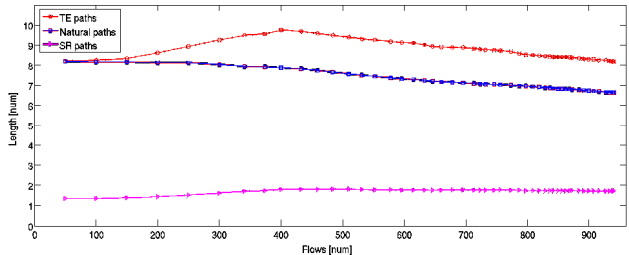

Figure 11 – Mean lengths of TE paths, natural paths, and SR paths vs. the number of allocated TE flows

We made a simple experimental evaluation of the processing time of the proposed TE/SR algorithms. Figure 12 reports the time spent for the computation of TE paths (flow assignment heuristic) and of SR paths (SR assignment algorithm). We use a PC with an Intel Core i7 2Ghz and 6GB RAM. Note that processing time of the flow assignment heuristic depends on the number of iterations of the optimization cycle, therefore a set of seemingly parallel lines can be appreciated in the figure (each one corresponds to a given number of cycles). The processing time of the SR assignment algorithm is negligible with respect to the flow assignment heuristic. In the considered range (up to 900 admitted flows) it was possible to run both algorithms and allocate the flows in less than 8 seconds. This performance seems adequate for periodic (e.g. nightly) reallocation procedures that aim to evenly redistribute the load on the network links.

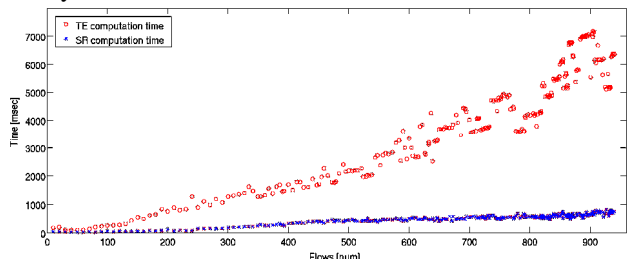

Figure 12 – Execution time of the algorithms

In the first class of experiments, we only evaluated the SR paths but did not try to actually allocate them. For the second class of experiments, a smaller testbed network composed by 12 IP/MPLS connected to a SDN controller has been considered (Figure 13). This network has been emulated on Mininet, in an example reported in [8] we deployed 16 IP VLLs on unidirectional SR paths, verifying the data plane connectivity among the CEs.

In the example shown in Figure 13, we deployed two bidirectional IP VLLs between the CE nodes A and B. IP VLL-1 requires 7.87 Gb/s (in both directions), IP VLL-2 requires 1.2 Gb/s (in both directions). All the links of the emulated topology have a nominal capacity of 10 Gb/s. Figure 13 shows the allocation provided by the TE flow assignment heuristic followed by the SR allocation algorithm. The IP VLL-1 is forwarded in the network with a SR path of length one. The IP VLL 2 in the A>B direction has an SR path of length 2, while in the B>A direction has a path of length 3.

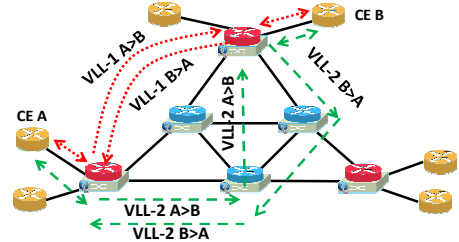

Figure 13 – Network topology with 12 nodes deployed on Mininet

VII. CONCLUSIONS

In this paper we presented a solution for Traffic Engineering with Segment Routing, based on MPLS forwarding. The solution has been implemented and tested on Mininet emulator and the code is available as open source. The solution is based on a known heuristic for TE and on a proposed SR path assignment algorithm. The initial performance analysis provides encouraging results concerning the feasibility of the approach.

ACKNOWLEDGMENTS

This work builds on the results of DREAMER project, partly funded by the EU as one of the beneficiary projects of the GÉANT Open Call research initiative.